\documentclass[12pt]{article}
\usepackage{amsmath, amssymb, amsthm, epsfig, natbib}
\usepackage{bbm,mathrsfs}
\usepackage{comment}
\usepackage{graphicx, graphics, rotating}
\usepackage{array}
\usepackage{multirow}
\usepackage{hyperref}
\usepackage{epstopdf}

\newtheorem{proposition}{Proposition}
\newtheorem{example}{Example}

\def\E{\mbox{\rm E}}

\def\argmin{\mbox{\rm argmin}}

\def\o{\mbox{\rm o}}

\renewcommand{\vec}[1]{\mbox{\boldmath ${#1}$}}
\def\vx{\mathbf{x}}

\def\vt{\mathbf{t}}

\def\ve{\mathbf{e}}

\def\vv{\mathbf{v}}

\def\vbeta{\vec{\beta}}
\def\vtheta{\vec{\theta}}

\def\vrho{\vec{\rho}}

\def\cD{\mathcal{D}}

\newcommand{\1}{\mathbf{1}}

\date{}

\begin{document}
\title{A Bayesian Stochastic Approximation Method}
\author{Jin Xu\footnote{Corresponding author: School of Statistics, East China Normal University, 500 Dongchuan Road, Shanghai 200241, China,
e-mail: {\sf{jxu@stat.ecnu.edu.cn}}}\quad and Cui Xiong and Rongji Mu
\\ School of Statistics\\
East China Normal University\\
Shanghai 200241, China}
\maketitle

\begin{abstract}
Motivated by the goal of improving the efficiency of small sample design, we propose a novel Bayesian stochastic approximation method to estimate the root of a regression function. The method features adaptive local modelling and nonrecursive iteration. Strong consistency of the Bayes estimator is obtained. Simulation studies show that our method is superior in finite-sample performance to Robbins--Monro type procedures. Extensions to searching for extrema and a version of generalized multivariate quantile are presented.
\end{abstract}

\vskip .5in


\section{Introduction}

We consider the problem of finding the unique root $\theta$ of a unknown function $M$ in the regression model
\begin{equation}\label{eq:reg-origin}
y_n=M(x_n)+\varepsilon_n,\quad n=1,2,\ldots
\end{equation}
where $\varepsilon_n$ is unobservable random error. The approach by stochastic approximation uses a sequential design strategy to successively choose $x_n$ on which the response $y_n$ is observed with mean $M(x_n)$ so that $x_n$ converges to $\theta$ in some sense. The feature of response-adaptiveness is attractive and can often be more efficient than fixed sample design. Over years, stochastic approximation and its variants have broad applications in design of experiments, clinical trials, dynamic programming, sequential learning, to name just a few \citep{Finney:78,Kushner:Yin:97,Spall:03}.

Here we give a brief review which is by no means to be complete but just covers some major progresses. In the fundamental paper of \citet{Robbins:Monro:51}, they proposed a recursive design of the form
\begin{equation}\label{eq:RM}
x_{n+1}=x_n-a_ny_n,
\end{equation}
where $a_n$ are positive constants, and showed that $x_n$ converges to $\theta$ in probability when $\sum_{n=1}^{\infty}a_n=\infty$ and $\sum_{n=1}^{\infty}a_n^2<\infty$ assuming $M$ satisfies some regularity conditions. It is a stochastic analogy to the deterministic Newton's method where $x_{n+1}=x_n-M(x_n)/M'(x_n)$ (The prime denotes the first derivative.) and is referred as Robbins--Monro procedure. The almost sure convergence was later proved through different approaches \citep{Dvoretzky:56,Gladyshev:65,Robbins:Siegmund:71}. Inspired by the Liapounov functions in the stability theory of ordinary differential equations, \citet{Sacks:58} established the asymptotic normality of $x_n$ and showed that under certain regularity conditions the asymptotically optimal choice of $a_n$ in (\ref{eq:RM}) is $a_n=(n\beta^*)^{-1}$ where $\beta^*=M'(\theta)$. (See also \citet{Chung:54},  \citet{Burkholder:56}, \citet{Hodges:Lehmann:56}.)

Ever since, much effects have been made to estimate $\beta^*$. \citet{Lai:Robbins:79,Lai:Robbins:81} proposed an adaptive Robbins--Monro procedure in the form of
\begin{equation}\label{eq:aRM}
x_{n+1}=x_n-(nb_n)^{-1}y_n,
\end{equation}
where $b_n$ is a truncated version of the least square estimate of the regression slope given by
$\widehat\beta_n=\sum_{i=1}^ny_i(x_i-\overline{x}_n)/\sum_{i=1}^n(x_i-\overline{x}_n)^2$ and
$\overline{x}_n=n^{-1}\sum_{i=1}^nx_i$.
Strong consistency of $b_n$ was established \citep{Lai:Robbins:81,Lai:Robbins:82}. We refer the readers to \citet{Venter:67}, \citet{Anderson:Taylor:76}, \citet{Anbar:78} and \citet{Anderson:Taylor:79} for some related versions. An excellent review about these variants is given by \citet{Lai:03}.

In a different route, \citet{Ruppert:88} and \citet{Polyak:Juditsky:92} proposed using averaged trajectories of (\ref{eq:RM}), $\overline{x}_n$, to estimate the root and demonstrated the almost sure convergence when $a_n$ satisfies the condition of being sufficiently slowly decreasing in the sense of $a_n\rightarrow 0$ and $(a_n-a_{n+1})/a_n=\o(a_n)$.

An important case of (\ref{eq:reg-origin}) is when $M$ is a distribution function and $y_n$ is binary response. Then, the Robbins--Monro procedure for finding the $\alpha$-quantile of $M$, assuming it is unique, is given by
\begin{equation}\label{eq:RM-b}
x_{n+1}=x_{n}-a_n(y_n-\alpha).
\end{equation}
The corresponding adaptive version is $x_{n+1}=x_{n}-(nb_n)^{-1}(y_n-\alpha)$.

The rationale of these procedures is clear. When observing a `success' at the $n$th step (such as explosion in the sensitivity experiment or occurrence of adverse events in dose-finding clinical trial), reduce the current level for the next design point; when observing a `failure', increase the current level for the next design point. As the number of iteration increases, the magnitude of change converges to zero. This type of scheme is in a similar spirit to the `up-and-down' method \citep{Dixon:Mood:48, Dixon:65} for estimating the median in sensitivity experiments. To estimate $\beta^*$ in the binary data case, \citet{Wu:85} proposed fitting a two-parameter logit model for the available data to obtain an initial maximum likelihood estimate (MLE) of $x_{n+1}$. Some initial runs are required to have the condition for the existence and uniqueness of this MLE met. (See also \citet{Sitter:Wu:93}.) An important contribution by \citet{Joseph:04} is the proposal of an efficient Robbins--Monro procedure which entails the recursion
\begin{equation}\label{eq:J04}
x_{n+1}=x_{n}-a_n(y_n-\alpha_n)
\end{equation}
where
\begin{equation*}\label{eq:an-bn}
a_n=\frac{\beta\tau_n^2}{\alpha_n(1-\alpha_n)(1+\beta^2\tau_n^2)^{1/2}}
\phi\left\{\frac{\Phi^{-1}(\alpha)}{(1+\beta^2\tau_n^2)^{1/2}}\right\}, \quad
\alpha_n=\Phi\left\{\frac{\Phi^{-1}(\alpha)}{(1+\beta^2\tau_n^2)^{1/2}}\right\},
\end{equation*}
$\tau_{n+1}^2=\tau_n^2-\alpha_n(1-\alpha_n)a_n^2$, $\beta=M'(\theta)/\phi(\Phi^{-1}(\alpha))$, $\Phi$ and $\phi$ are the distribution function and density of the standard normal variable respectively. The introduction of constant sequence $\alpha_n\rightarrow\alpha$ helps reduce the oscillation of $x_n$ at early steps. It is shown to have a faster convergence than the usual Robbins--Monro procedure when $\alpha$ takes extreme values. \citet{Wu:Tian:14} proposed a three-phase design that combines some initial design and Joseph's efficient modification to obtain a more steady method. Recently, \citet{Toulis:Airoldi:15} proposed an implicit stochastic approximation method which improves the classic Robbins--Monro procedure by a stochastic fixed-point equation. It requires to run many additional experiments at every step of (\ref{eq:RM}). Thus, it may not be feasible for a small sample design.

Other model-based designs for quantal response focus on estimation of the coefficients of a parametric model~\citep{Wu:86,Chaloner:Larntz:89,Chaudhuri:Mykland:93,Neyer:94, Dror:Steinberg:06,Dror:Steinberg:08, Hung:Joseph:14}. The advantage of this approach is that one can use a single design to estimate the global response curve that includes all quantiles. While the disadvantages are that i) it needs to make assumptions (about the model and/or hyperparameters); and ii) the designs usually require initial data to start with which can be as many as ten or more. \citet{Hung:Joseph:14} proposed a simple Bayesian version of \citet{Wu:85}'s logit-MLE method, which makes the design fully sequential from $n=1$. It postulates independent informative priors on the parameters of a logistic model for $M$ given by $F(x)=[1+\exp\{-(x-\mu)/\sigma\}]^{-1}$ with $\mu\sim N(\mu_0,\tau^2)$ and $\sigma\sim\exp(\xi)$. And the sequential design estimates the $\alpha$-quantile by $x_{n+1}=\hat\mu_n+\hat\sigma_n\log\{p/(1-p)\}$, where $\hat\mu_n$ and $\hat\sigma_n$ are the maximum-a-posteriori (MAP) estimate of $(\mu,\sigma)$ after $n$ samples.

In this paper, we limit our study to the root finding problem. We point out several limitations associated with the Robbins--Monro type procedures. First, for these algorithm-based procedures such as (\ref{eq:RM}), the averaged trajectory of (\ref{eq:RM}) and (\ref{eq:J04}), the adaptation through the last experiment data $(x_n, y_n)$ via recursion is subject to inadequacy. Experiments at points in a neighborhood would carry useful information for $\theta$ as well especially in the early stage. Second, large oscillation caused by these up-or-down recursions in early iteration can be harmful and inefficient. Third, for the procedures such as (\ref{eq:aRM}) that reply heavily on the estimation of $\beta^*$, as $x_n$ clusters around to $\theta$, little information is gained to estimate $\beta^*$ directly. So even for consistent estimator, the finite-sample performance can still be far from satisfaction from a practical point of view.

On the other hand, the Bayesian paradigm is known to be suitable for such adaptive learning problem. Some applications in a closely related problem of dose-finding in clinical trials have been reported \citep{Cheung:10,Thall:10}. Like \citet{Hung:Joseph:14}'s method, Bayesian models are used to update the underlying distribution \textit{globally}. Little has been seen for solving the \textit{local} root for $\alpha$-quantile directly. Using martingale theory, \citet{Hu:98} established the strong consistency of the Bayes estimator under a general setting of a nonlinear regression model. We will make use of this result for later development.

Motivated by the aforementioned drawbacks of the Robbins--Monro type methods and the advantage of Bayesian approach, we propose a novel model-based stochastic approximation procedure that circumvents direct estimation of $\beta^*$ through integration. Specifically, the new method builds a local linear model for $M$ around $x_n$ and obtains the Bayes estimator as a nonrecursive solution for $x_{n+1}$. Strong consistency is obtained. These constitute the main contents of Section~\ref{sec:method}. In Section~\ref{sec:remark}, we give a few important remarks and insights of the proposed method that lead to more efficient algorithm. More importantly, in Section~\ref{sec:simu} we demonstrate by simulation that the proposed method yields a smooth search path and results in a superior finite-sample performance to the competing methods. In Section~\ref{sec:application}, we present applications of the new method to the general root-finding problem in (\ref{eq:reg-origin}) and Kiefer--Wolfowitz procedure \citep{Kiefer:Wolfowitz:52} to find the minimum of an unknown function. In Section~\ref{sec:extension}, we extend the proposed method to estimate a version of generalized multivariate quantile. Section~\ref{sec:disc} concludes the paper with some discussions.

\section{Method}\label{sec:method}
We begin with the problem of quantile estimation with binary responses under the setting in (\ref{eq:RM-b}).

First, we introduce two preliminary processes before sequential experiment. (i) Scale the search domain of $x$ to the interval $(0,1)$. It can be done easily once we have some general idea of the range of $x$. (ii) Divide the interval $(0,1)$ equally into $s$ subintervals. We will provide guideline for the selection of $s$ in Section~\ref{sec:choice-s}.

Denote the (scaled) data up to the $n$th step by $\cD_n=\{(x_i,y_i): i=1,\ldots,n\}$. Next, we construct a local Bayesian model based on the current point $x_n$. Observe that $x_n$ is contained in the subinterval $(v_0,v_1)$, where $v_0=(\lceil x_ns\rceil-1)/s$, $v_1=\lceil x_ns\rceil/s$, and $\lceil\cdot\rceil$ is the ceiling function. Approximate $M(x)$ in $(v_0, v_1)$ by the segment of a line through the point $(\theta, \alpha)$ with positive slope $\beta$ given by
\begin{equation}\label{eq:Fx}
F(x)=\alpha+\beta(x-\theta),\quad x\in (v_0, v_1).
\end{equation}
Note that $\theta$ itself is not necessarily in $(v_0,v_1)$.

For the convenience of later calculation, denote $\widetilde\beta=\beta(v_1-v_0)\  (=\beta/s)$.
Let
\begin{equation}\label{eq:rho-in-beta}
\rho_0=F(v_0)=\alpha+s\widetilde\beta(v_0-\theta) \quad\textrm{and}\quad
\rho_1=F(v_1)=\alpha+s\widetilde\beta(v_1-\theta).
\end{equation}
Then, $\widetilde\beta$ and $\theta$ are 1-1 connected with $\rho_0$ and $\rho_1$ through
\begin{equation}\label{eq:theta-in-rho}
\widetilde\beta=\rho_1-\rho_0 \quad\textrm{and}\quad
\theta=\frac{\rho_1- \alpha}{\rho_1- \rho_0}v_0+\frac{\alpha-\rho_0}
{\rho_1-\rho_0}v_1.
\end{equation}

Assume that the joint prior of $(\rho_0, \rho_1)$ is uniform with density
\begin{equation}\label{eq:prior-rho}
h(\rho_0,\rho_1)=\frac{2I(\rho_L<\rho_0<\rho_1<\rho_U)}{(\rho_U-\rho_L)^2},
\end{equation}
where $0\le\rho_L<\alpha<\rho_U\le 1$ are two given constants, and $I(\cdot)$ is the indicator function. For example, the constants $\rho_L=0$ and $\rho_U=1$ are considered to be noninformative. We have more discussion about the determination of $\rho_L$ and $\rho_U$ in Section~\ref{sec:post-rho}. It should be noted that under this prior, $\theta$ can take value in $(-\infty,\infty)$ through (\ref{eq:theta-in-rho}) as linear extrapolation. For later development, we will restrict the calculation of the posterior distribution of $\theta$ in the domain $(0,1)$ by truncation. And we will introduce other prior which meets the restriction for $\theta\in(0,1)$ in Section~\ref{sec:prior-rho}.

The subsequent development for finding the posterior distribution of $\theta$ is standard. After accounting for the Jacobian from (\ref{eq:rho-in-beta}), the joint prior density of $(\theta,\widetilde\beta)$ is
\begin{equation*}\label{eq:jpdf-beta-theta}
h(\theta,\widetilde\beta)=\frac{2s\widetilde\beta
I\left(\rho_L<\alpha+s\widetilde\beta(v_0-\theta)
<\alpha+s\widetilde\beta(v_1-\theta)<\rho_U\right)}{(\rho_U-\rho_L)^2},
\end{equation*}
which can be expressed as
\begin{equation}\label{eq:jpdf-beta-theta-2}
h(\theta,\widetilde\beta)=\frac{2s\widetilde\beta
I\left(0<\widetilde\beta<\eta(\theta)\right)}{(\rho_U-\rho_L)^2}
\end{equation}
with
\begin{equation}\label{eq:eta}
\eta(\theta)=\frac{(\rho_U-\alpha)I(\theta\le\theta_0)}{s(v_1-\theta)}+
\frac{(\alpha-\rho_L)I(\theta>\theta_0)}{s(\theta-v_0)}, \quad \theta_0=\frac{(\rho_U-\alpha)v_0+(\alpha-\rho_L)v_1}{\rho_U-\rho_L}.
\end{equation}
Note that $0<\eta(\theta)<\rho_U-\rho_L$. Integrating out $\widetilde\beta$ in (\ref{eq:jpdf-beta-theta-2}) and imposing the restriction that $0<\theta<1$, we obtain the prior density of $\theta$ as
\begin{equation*}\label{eq:h0}
h_0(\theta)=\frac{s\eta^2(\theta)}{c_0(\rho_U-\rho_L)^2},
\end{equation*}
where $c_0=\int_0^1s\eta^2(\theta)/(\rho_U-\rho_L)^2d\theta$ is the normalization constant. 

Next, we will only use the design points contained in $(v_0,v_1)$ to update the Bayesian model. This idea of using most recent design points is also seen in \citet{Anbar:78} to estimate $\beta^*$.

Denote the subsequence of $x_n$ in $(v_0,v_1)$ by $x_{i_1},\ldots,x_{i_m}$. Clearly, $1\le m\le n$ since at least $x_n$ is in $(v_0,v_1)$. Denote the likelihood function of $(\theta,\widetilde\beta)$ at point $(x_i, y_i)$ by $L_i$, which is expressed as
\begin{equation}\label{eq:Li}
L_i(\theta,\widetilde\beta)=F(x_i)^{y_i}\{1-F(x_i)\}^{1-y_i}=a_i+b_i(\theta)\widetilde\beta,
\end{equation}
where
\begin{equation}\label{eq:ai-bi}
a_i=\alpha^{y_i}(1-\alpha)^{1-y_i}=1-y_i+(2y_i-1)\alpha,\quad b_i(\theta)=s(2y_i-1)(x_i-\theta).
\end{equation}
By (\ref{eq:jpdf-beta-theta-2}) and (\ref{eq:Li}), the posterior distribution of $(\theta,\widetilde\beta)$ is proportional to
\begin{equation}\label{eq:post-beta-theta}
h(\theta,\widetilde\beta)\prod_{j=1}^mL_{i_j}(\theta,\widetilde\beta)=
\frac{2s\widetilde\beta I\{0<\widetilde\beta<\eta(\theta)\}}{(\rho_U-\rho_L)^2}
\prod_{j=1}^m\{a_{i_j}+b_{i_j}(\theta)\widetilde\beta\}.
\end{equation}
For $r=0,1,\ldots,m$, express the coefficient of $\widetilde\beta^r$ in $\prod_{j=1}^m\{a_{i_j}+b_{i_j}(\theta)\widetilde\beta\}$ as
\begin{equation}\label{eq:dm-r}
d_{m,r}(\theta)=\sum_{B\in\Omega_{m,r}}\prod_{t\in B^c}a_{i_t}\prod_{k\in B}b_{i_k}(\theta),
\end{equation}
where $\Omega_{m,r}$ is the collection of $m$-choose-$r$ distinct subsets of $r$ indices out of $\{1,\ldots,m\}$ and $B^c=\{1,\ldots,m\}\backslash B$. We emphasize that $d_{m,r}(\theta)$ only depends on data observed in the subinterval.

Integrating out $\widetilde\beta$ in (\ref{eq:post-beta-theta}), we get the posterior distribution of $\theta$ as
\begin{equation}\label{eq:hm}
h_m(\theta)=
\frac{2s}{c_m(\rho_U-\rho_L)^2}
\sum_{r=0}^m \frac{d_{m,r}(\theta)\eta^{r+2}(\theta)}{r+2}
=\frac{2c_0h_0(\theta)}{c_m}
\sum_{r=0}^m \frac{d_{m,r}(\theta)\eta^{r}(\theta)}{r+2},
\end{equation}
where $c_m$ is the normalization constant. A few points are worthy of being noted. First, $h_m(\theta)$ is a two-piecewise homogeneous polynomial of order $-2$ and is differentiable everywhere except at $\theta_0$. Second, $h_m(\theta)$ is invariant to the permutation of the points in the subsequence. Third, the modification of $h_m$ to the prior $h_0$ takes place in a multiplicative fashion. The weighted summand $d_{m,r}(\theta)\eta^{r}(\theta)$ can be viewed as the $r$th order interaction of the points in the subinterval. Moreover, we can write $d_{m,r}(\theta)$ recursively as
\begin{equation}\label{eq:dm-r-recursive}
d_{m,r}(\theta)=d_{m-1,r}(\theta)a_{i_m}+d_{m-1,r-1}(\theta)b_{i_m}(\theta), \quad r=0,\ldots,m,
\end{equation}
where $d_{0,0}=1$, $d_{m-1,-1}=d_{m-1,m}=0$. It provides a simple way to obtain $d_{m,r}$ successively. Based on (\ref{eq:dm-r-recursive}), we express $c_mh_m(\theta)$ in a recursive form as
\begin{equation}\label{eq:hm-recursive}
c_mh_m(\theta)=c_{m-1}h_{m-1}(\theta)
\left\{a_{i_m}+b_{i_m}(\theta)\eta(\theta)R_{m-1}(\theta)\right\},
\end{equation}
where $R_{m-1}(\theta)=\sum_{r=0}^{m-1}(r+3)^{-1}d_{m-1,r}(\theta)\eta^r(\theta)/
\sum_{r=0}^{m-1}(r+2)^{-1}d_{m-1,r}(\theta)\eta^r(\theta)$.

We summarize the above results in the following proposition.
\begin{proposition}\label{prop:post-hm}
Assume that the joint prior of $(\rho_0,\rho_1)$ associated with the subinterval $(v_0,v_1)$ is uniform with density (\ref{eq:prior-rho}). Then, the posterior distribution of $\theta$ restricted in $(0,1)$ is given in (\ref{eq:hm}) satisfying a recursion in (\ref{eq:hm-recursive}).
\end{proposition}

Finally, we set the next point to be the Bayes estimator with respect to $h_m$, i.e.
\begin{equation}\label{eq:post-mean}
x_{n+1}=\E_{h_m}(\theta).
\end{equation}
Since $h_m(\theta)$ or $c_mh_m(\theta)$ is completely determined in (\ref{eq:hm}), $x_{n+1}$ can be easily calculated up to a desired precision. We can also easily obtain an equal tail credible interval for $\theta$ based on $h_m$.

When $M$ is linear as $F$ in (\ref{eq:Fx}), it is clear that the random error for the binary response $y_n$ satisfies the conditions $\E(\varepsilon_n\mid\varepsilon_1,\ldots,\varepsilon_{n-1})=0$ and $\E(\varepsilon_n^2)<\infty$. Then, by Theorem~1 of \citet{Hu:98}, we have the following result about the consistency of the procedure.
\begin{proposition}
For binary response with mean value given by the model (\ref{eq:Fx}), the Bayesian stochastic approximation procedure given by (\ref{eq:post-mean}) is strongly consistent.
\end{proposition}

When $M$ is nonlinear, by Taylor expansion $M(x)$ differs from $F(x)$ by a quantity bounded by $\sup_{x\in(v_0,v_1)}|M''(x)|/(2s^2)$, where $M''$ denotes the second derivative assuming it exists. As $n$ increases, we can increase $s$ so that the local linear approximation is well maintained. Thus, we expect the consistency of the procedure to hold. We demonstrate its superb finite-sample performance in Section~\ref{sec:simu}.

\begin{figure}
\centerline{\epsfig{file=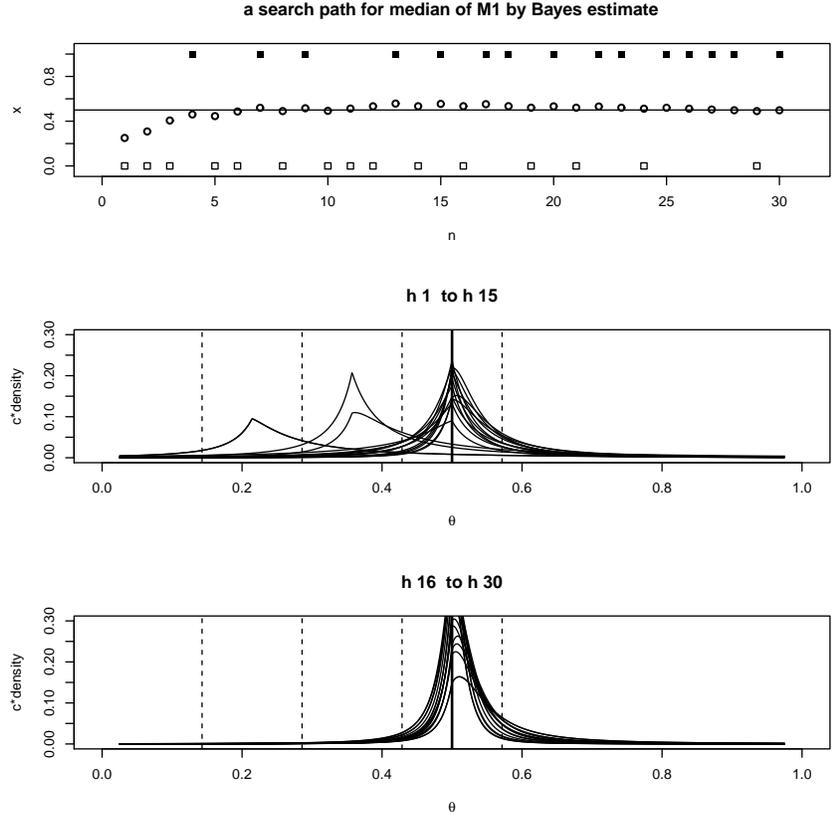, width=4.5in}}\par
\caption{One search path up to 30 steps for the median of $M_1(x)$ and the evolution of the corresponding posterior densities $h_{(n)}$ (which equals $h_m$ for some $m$ in the associated subinterval) obtained by the proposed method using the Bayes estimate. In the upper panel, the observed values of $y_n$ are depicted along the x-axis by empty squares for 0 and filled squares for 1. In the middle and lower panels, the y-axes are the corresponding $h_m$ multiplied by $c_m/\prod_{j=1}^ma_{i_j}$ for illustration purpose. The dotted lines indicate the endpoints of the subintervals and the solid line indicates the root 0.5.}\label{fig:hmChangeBayes}
\end{figure}

\begin{figure}
\centerline{\epsfig{file=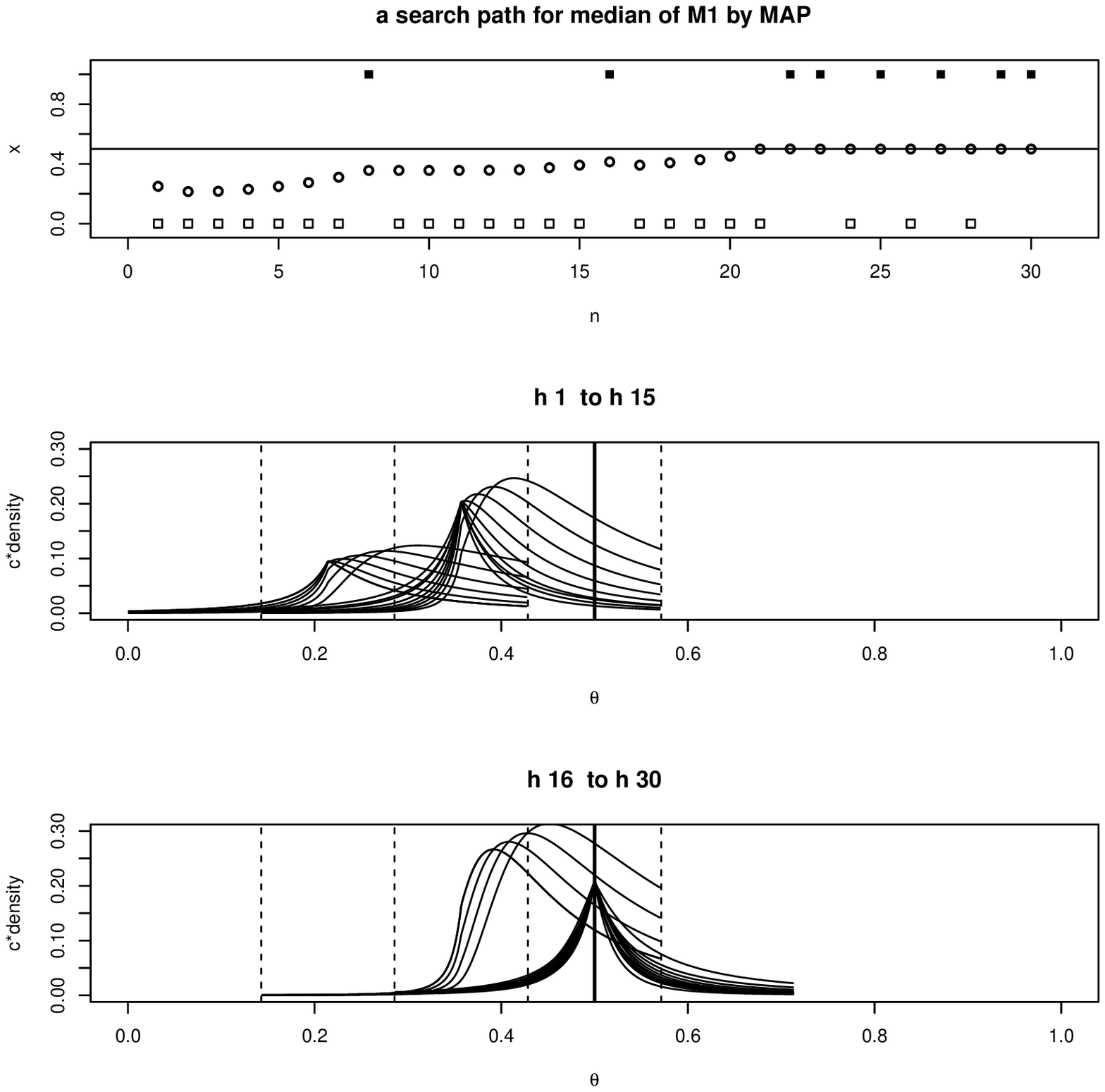, width=4.5in}}\par
\caption{One search path up to 30 steps for the median of $M_1(x)$ and the evolution of the corresponding posterior densities $h_{(n)}$ (which equals $h_m$ for some $m$ in the associated subinterval) obtained by the proposed method using the MAP estimate. In the upper panel, the observed values of $y_n$ are depicted along the x-axis by empty squares for 0 and filled squares for 1. In the middle and lower panels, the y-axes are the corresponding $h_m$ multiplied by $c_m/\prod_{j=1}^ma_{i_j}$ for illustration purpose. The dotted lines indicate the endpoints of the subintervals and the solid line indicates the root 0.5.}\label{fig:hmChangeMAP}
\end{figure}

Beside the Bayes estimator, we can also use the posterior mode, i.e. maximum a posterior  (MAP) estimator, for the next point. We illustrate the procedure by an example.
\begin{example}\label{exp:BayesvsMAP}
Let $M_1(x)=\Phi(6x-3)$ for $x\in(0,1)$. Consider estimating the median of $M_1$. Set $x_1=0.25$ and $s=7$. And set $\rho_L=0$ and $\rho_U=1$ for all subintervals. Figures~\ref{fig:hmChangeBayes} and \ref{fig:hmChangeMAP} demonstrate one search path up to 30 steps and the evolution of the corresponding posterior distributions $h_{(n)}$ (which equals $h_m$ for some $m$ in the associated subinterval) obtained by the proposed method using the Bayes estimate and the MAP estimate, respectively. Notice that $c_m/(c_{m-1}a_{i_m})\rightarrow 1$. For the purpose of illustrating the shape of $h_m$, we multiply $h_m$ by $c_m/\prod_{j=1}^ma_{i_j}$ to make the amplified $h_m$s in a comparable scale. It is seen that both sequences move across three subintervals and gradually converge to the median 0.5. The Bayes estimate appears to converge faster than the MAP estimate as it is more aggressive to move across a subinterval. While, the MAP estimate tends to yield a conservative movement and a more smooth path. These patterns are consistent to the properties of mean and median with respect to the skewness of a distribution.
\end{example}

\section{Remarks}\label{sec:remark}
In this subsection, we give a few important remarks and insights of the proposed method that can lead to more efficient algorithm.

\subsection{Posterior distributions of $\rho_0$ and $\rho_1$}\label{sec:post-rho}
By (\ref{eq:Fx}) and (\ref{eq:rho-in-beta}), we have linear interpolation for $x_i\in (v_0,v_1)$ as $F(x_i)=q_i\rho_0+(1-q_i)\rho_1$ with $q_i=(v_1-x_i)/(v_1-v_0)$. Express the individual likelihood in (\ref{eq:Li}) in terms of $(\rho_0,\rho_1)$ as
\[L_i(\rho_0,\rho_1)=1-y_i+(2y_i-1)q_i\rho_0+(2y_i-1)(1-q_i)\rho_1.\]
Then, following the same routine as in Section~\ref{sec:method} for $\theta$, we obtain the marginal posterior distributions of $\rho_0$ and $\rho_1$ as follows.
\begin{proposition}\label{prop:post-rho}
Assume that the joint prior of $(\rho_0,\rho_1)$ associated with the subinterval $(v_0,v_1)$ is uniform with density (\ref{eq:prior-rho}). Then, the posterior distribution of $\rho_0$ is
\begin{equation*}\label{eq:hm-rho-0}
h_m(\rho_0)=
\frac{2}{c_m^*(\rho_U-\rho_L)^2}
\sum_{r=0}^m \frac{d_{m,r}(\rho_0) (\rho_U^{r+1}-\rho_0^{r+1})}{r+1},
\end{equation*}
where $d_{m,r}$ is defined in the same form as (\ref{eq:dm-r}) with
\[a_i=1-y_i+(2y_i-1)q_i\rho_0,\quad b_i=(2y_i-1)(1-q_i),\]
and $c_m^*$ is the normalization constant. And the posterior distribution of $\rho_1$ is
\begin{equation*}\label{eq:hm-rho-1}
h_m(\rho_1)=
\frac{2}{c_m^{**}(\rho_U-\rho_L)^2}
\sum_{r=0}^m \frac{d_{m,r}(\rho_1) (\rho_1^{r+1}-\rho_L^{r+1})}{r+1},
\end{equation*}
where $d_{m,r}$ is defined in the same form as (\ref{eq:dm-r}) with
\[a_i=1-y_i+(2y_i-1)(1-q_i)\rho_1,\quad b_i=(2y_i-1)q_i,\]
and $c_m^{**}$ is the normalization constant.
\end{proposition}
The recursion in (\ref{eq:dm-r-recursive}) also holds for $d_{m,r}(\rho_0)$ and $d_{m,r}(\rho_1)$. Like $h_m(\theta)$ in Section~\ref{sec:method}, $h_m(\rho_0)$ and $h_m(\rho_1)$ are completely determined given the data.

When $x_n$ enters a subinterval for either the first time or re-visit, we can use the posterior distributions obtained from the previous subinterval to update $\rho_L$ or $\rho_U$ for the uniform prior of the current subinterval. More specifically, suppose that $x_n$ moves forward from the $t$th subinterval to the $(t+1)$th subinterval. Then we can set the fifth percentile of the posterior distribution of $\rho_1$ of the $t$th subinterval as $\rho_L$ for the $(t+1)$th subinterval. And suppose that $x_n$ moves downward from the $t$th subinterval to the $(t-1)$th subinterval. Then we can set the 95th percentile of the posterior distribution of $\rho_0$ of the $t$th subinterval as $\rho_U$ for the $(t-1)$th subinterval. In this way, the information from the neighboring subinterval is used for the new local model. We will use this strategy in the subsequent numerical study. As seen in simulation, these lower or upper fifth percentile can actually narrow the range of the uniform prior significantly as data cumulates.

\subsection{Posterior distribution of $\widetilde\beta$}\label{sec:remark-post-beta}
The joint prior $h(\theta,\widetilde\beta)$ in (\ref{eq:jpdf-beta-theta}) can also be written as
\begin{equation*}\label{eq:jpdf-beta-theta-1}
h(\widetilde\beta,\theta)=\frac{2s\widetilde\beta
I\left\{0<\widetilde\beta<\rho_U-\rho_L,\ \ell(\widetilde\beta)<\theta<u(\widetilde\beta)\right\}}{(\rho_U-\rho_L)^2},
\end{equation*}
with
\begin{equation*}\label{eq:L-U}
\ell(\widetilde\beta)=v_1-\frac{\rho_U-\alpha}{s\widetilde\beta},\quad u(\widetilde\beta)=v_0+\frac{\alpha-\rho_L}{s\widetilde\beta},
\end{equation*}
which indicates that $\theta$ given $\widetilde\beta$ is uniform. Note that without further restriction of $\widetilde\beta$, the interval $(\ell(\widetilde\beta),u(\widetilde\beta))$ can be as wide as $(-\infty,\infty)$ as pointed out before. To impose the conditions $\ell(\widetilde\beta)\ge 0$ and $u(\widetilde\beta)\le 1$ requires $\widetilde\beta\ge \widetilde\beta_0$ where $\widetilde\beta_0=\max\{\frac{\rho_U-\alpha}{sv_1},\frac{\alpha-\rho_L}{s(1-v_0)}\}$. Then, the marginal prior of $\widetilde\beta$ is
\[g_0(\widetilde\beta)=\frac{2(\rho_U-\rho_L-\widetilde\beta)}{\widetilde{c}_0(\rho_U-\rho_L)^2},\]
where $\widetilde{c}_0$ is the normalization constant (over $(\widetilde{\beta}_0,\rho_U-\rho_L)$).

Secondly, express $L_i(\theta,\widetilde\beta)$ in (\ref{eq:Li}) as
\begin{equation*}\label{eq:Li-2}
L_i(\theta,\widetilde\beta)=\widetilde{a}_i(\widetilde\beta)+\widetilde{b}_i(\widetilde\beta)\theta,
\end{equation*}
where
\begin{equation}\label{eq:ai-bi-2}
\widetilde{a}_i(\widetilde\beta)=1-y_i+(2y_i-1)(\alpha+s\widetilde\beta x_i),\quad \widetilde{b}_i(\widetilde\beta)=-(2y_i-1)s\widetilde\beta.
\end{equation}
Following the same steps in (\ref{eq:dm-r}) and (\ref{eq:hm}), we get
\begin{proposition}\label{prop:post-gm}
Assume that the joint prior of $(\rho_0,\rho_1)$ associated with the subinterval $(v_0,v_1)$ is uniform with density (\ref{eq:prior-rho}). Then, the posterior distribution of $\widetilde\beta$ is
\begin{equation*}\label{eq:gm}
g_m(\widetilde\beta)=
\frac{2s\widetilde\beta}{\widetilde{c}_m}
\sum_{r=0}^m \frac{d_{m,r}(\widetilde\beta) \left\{u(\widetilde\beta)^{r+1}-\ell(\widetilde\beta)^{r+1}\right\}}{r+1},
\end{equation*}
where $d_{m,r}$ is defined in the same form as (\ref{eq:dm-r}) with $a_i$ and $b_i$ replaced by $\widetilde{a}_i$ and $\widetilde{b}_i$ in (\ref{eq:ai-bi-2}) respectively, and $\widetilde{c}_m$ is the normalization constant.
\end{proposition}

\subsection{Investigation of $x_2$}
We present a detailed investigation of $x_2$ to reveal some features of the proposed procedure.

By (\ref{eq:hm}), we have
\[
h_1(\theta)=\frac{2s}{c_1}
\left\{\frac{a_1\eta^2(\theta)}{2}+\frac{b_1(\theta)\eta^3(\theta)}{3}\right\}.
\]
For simplicity, fix $\rho_L=0$ and $\rho_U=1$ in (\ref{eq:eta}) for $\eta$.

To examine the connection between $x_2$ and $x_1$, we first consider the MAP estimate for $x_2$. By solving $h_1'(\theta)=0$ and checking the sign of $h_1'(\theta)$ for cases of $\theta<\theta_0$ and $\theta>\theta_0$ where $\theta_0$ is defined in (\ref{eq:eta}), we obtain that
\begin{equation}\label{eq:x2}
x_2=
\left\{
\begin{array}{ll}
x_1-\frac{1-4\alpha}{2+\alpha}(v_1-x_1)=\theta_0-\frac{3(1-\alpha)}{2+\alpha}(t_0-x_1),
& \textrm{if}\ x_1<t_0\ \textrm{and}\ y_1=1, \\
x_1+\frac{4\alpha-3}{3-\alpha}(x_1-v_0)=\theta_0+\frac{3\alpha}{3-\alpha}(x_1-t_1),
& \textrm{if}\ x_1>t_1\ \textrm{and}\ y_1=0, \\
\theta_0,& \textrm{otherwise},
\end{array}
\right.
\end{equation}
where $t_0=3^{-1}(2+\alpha)v_0+3^{-1}(1-\alpha)v_1$ and $t_1=3^{-1}\alpha v_0+(1-3^{-1}\alpha)v_1$ which divide $(v_0, v_1)$ into subintervals $(v_0, t_0)$, $(t_0, t_1)$ and $(t_1, v_1)$ with fractions of $(1-\alpha)/3$, $2/3$ and $\alpha/3$, respectively.
And $\theta_0$ falls in these subintervals depending on $\alpha$ value in $(0,1/4)$, $[1/4,3/4]$, $(3/4,1)$ respectively.

A few interesting properties of the MAP estimate can be seen from (\ref{eq:x2}). First, when $t_0<x_1<t_1$, $x_2=\theta_0$ no matter $y_1=1$ or 0. This outcome enables the search path to possibly remain unchanged (with $1/4\le \alpha \le 3/4$) when the evidence of moving is not convincing. Second, when $1/4\le \alpha \le 3/4$, the values of $x_2$ under the first two situations of (\ref{eq:x2}) are rather counterintuitive. For instance, when $y_1=1$ with $x_1<t_0$, we have $x_1<x_2$. It would have been $x_1>x_2$ by Robbins--Monro type procedure. However, the procedure does yield  $x_2<\theta_0$. Similarly, when $y_1=0$ with $x_1>t_1$, we get $\theta_0<x_2<x_1$, which would have been $x_2>x_1$ by Robbins--Monro type procedure. This seemingly irrational move can actually avoid unnecessary oscillation of the search points in the absence of enough evidence and lead to a smooth path as seen in Figures~\ref{fig:hmChangeBayes} and \ref{fig:hmChangeMAP} in contrast to a zig-zag path in Robbins--Monro type procedure. Third, $x_2$ can take value outside $(v_0, v_1)$. For example, when $\alpha<1/4$, $x_1<t_0$ and $y_1=1$, we get $x_2<v_0$; and when $\alpha>3/4$, $x_1>t_1$ and $y_1=0$, we get $x_2>v_1$. It results in the search point moving into the neighboring subinterval and consequently starting a new local Bayesian model.

The explicit expression of the MAP for $x_3$ can also be derived based on $h_2$. It depends on $(x_1,y_1)$ and $(x_2,y_2)$ and is very complicated.


Next, by straightforward calculation, the Bayes estimate for $x_2$ is obtained as
\begin{align*}
&\frac{a_1\alpha^2}{c_1s}\left\{\log \frac{(1-v_0)s}{\alpha} - \frac{v_0(\theta_0-1)s}{(1-v_0)\alpha}\right\}+
\frac{a_1(1-\alpha)^2}{c_1s}\left\{-\log\frac{v_1s}{1-\alpha}+\frac{\theta_0s}{1-\alpha}\right\}\\
+&\frac{2(2y_1-1)\alpha^3}{3c_1s}\left\{-\log \frac{(1-v_0)s}{\alpha} + \frac{(x_1-2v_0)(1-\theta_0)s}{(1-v_0)\alpha}+
\frac{v_0(x_1-v_0)(1-\theta_0)(\theta_0+1-2v_0)s^2}{2(1-v_0)^2\alpha^2}\right\} \\
+&\frac{2(2y_1-1)(1-\alpha)^3}{3c_1s}\left\{-\log\frac{v_1s}{1-\alpha}
+\frac{(2v_1-x_1)\theta_0s}{v_1(1-\alpha)}- \frac{\theta_0(v_1-x_1)(2v_1-\theta_0)s^2}{2v_1(1-\alpha)^2}\right\}.
\end{align*}
We can hardly interpret the connection of $x_2$ with $x_1$ from this analytic expression except that $x_2$ is a linear function of $x_1$. However numerical analysis shows that $x_2$ also processes similar features as those described for the MAP estimate.

At last, inspired by the above investigation of $x_2$, we find the proposed procedure is conservative in the sense of moving in large steps. So instead of choosing $x_1$ arbitrarily, we set $x_1=0.5$, the middle of the search domain, as the starting point to begin cumulating information.

\subsection{Choice of $s$}\label{sec:choice-s}
The number of subintervals $s$ determines the size of the neighborhood up on which a local model is built. When $\alpha$ is around the middle range, say $0.4\sim 0.6$, an integer in the range of $3\sim 10$ can usually yield a quick convergence in a moderate number of iterations.
When $\alpha$ is close to extreme values, implying rare event of `success' or `failure' in experiment, we wish the search sequence to be conservative in moving in small steps especially in the early iterations. Therefore, a moderately large value of $s$ is recommended, say 20. And for the same reason, we recommend using MAP estimator instead of the Bayes estimator.

Second, to get a more efficient approximation and faster convergence, we recommend a two-stage procedure. That is to set $s$ to be a small number to quickly reach the vicinity of the target and then increase $s$ to a larger number for refined approximation. We provide a guideline for the choice of $s$ in Table~\ref{tab:s}. The odd numbers are chosen to avoid possible invalid denominators in $\eta$ in (\ref{eq:eta}) during numerical calculation.

\begin{table}[t]
\caption{Choice of $s$ for a two-stage procedure}\par
\label{tab:s}\par
\vskip .2cm
\centerline{\tabcolsep=5truept\begin{tabular}{clccc} \hline
 & $\alpha\in$ &$(0.4,0.6)$ & $(0.1,0.4)\cup (0.6,0.9)$ & $(0,0.1)\cup (0.9,1)$ \\
\hline
$n=1,\ldots,10$  &&5 & 9 & 13 \\
$n\ge 11$  && 9 & 17 & 23 \\
\hline
\end{tabular}}
\end{table}

Third, if during the search updated information about the range of $\theta$ becomes available, one can re-define the search domain and use the available data after rescaling.

\subsection{Alternative choice of prior $h(\rho_0,\rho_1)$}\label{sec:prior-rho}
As seen in Section~\ref{sec:method}, the uniform prior of (\ref{eq:prior-rho}) leads to simple calculation for the derivation, but induces distribution of $\theta$ outside $(0,1)$. Alternatively, one can use other prior, such as
\[h(\rho_0,\rho_1)=\frac{2v_1}{v_0}I\left(0<\rho_0<\frac{v_0\rho_1}{v_1}, 0<\rho_1<1\right),  \quad v_0\ne 0,\]
or even more informative prior to warrant $\theta\in(0,1)$. Then simple or explicit form of the posterior distribution may not be available. In this case, we can resort to Markov chain Monte Carlo method, e.g. Gibbs sampling, to obtain the empirical posterior distribution of $\theta$ after (\ref{eq:theta-in-rho}) and (\ref{eq:post-beta-theta}). However, because of scarcity of the data and simulation error, preliminary numerical study shows that resulting estimates are not as precise as those based on the exact distribution.

\section{Numerical comparisons}\label{sec:simu}

We compare the proposed Bayesian stochastic approximation method using Bayes estimator (denoted by BSA-Bayes) and MAP estimator (denoted by BSA-MAP) with the classic Robbins--Monro procedure in (\ref{eq:RM-b}) (denoted by RM), the efficient Robbins--Monro procedure in (\ref{eq:J04}) (denoted by RMJ), the averaged trajectory method by \citet{Ruppert:88} and \citet{Polyak:Juditsky:92} (denoted by RPJ), and the Bayesian version of Wu's logit-MLE method by \citet{Hung:Joseph:14} (denoted by Wu-MAP).

Consider the following six functions adopted from \citet{Joseph:04},
\[
\begin{array}{lcl}
M_2(x)=\Phi(\Phi^{-1}(\alpha)+x), &&
M_3(x)=\min(1,\max(0,\alpha+\frac{z}{3})), \\
M_4(x)=(1+\frac{1-\alpha}{\alpha}e^{-x})^{-1}, &&
M_5(x)=1-\exp\{\log(1-\alpha)e^x\}, \\
M_6(x)=(1+\frac{1-\surd{\alpha}}{\surd{\alpha}}e^{-x})^{-2}, &&
M_7(x)=\frac{1}{2}+\frac{1}{\pi}\tan^{-1}[x+\tan\{\pi(\alpha-\frac{1}{2})\}],
\end{array}
\]
which represent a shifted version of normal, uniform, logistic, extreme value, skewed logistic, and Cauchy distributions respectively with a common root at zero for all $\alpha$-quantiles.

Since all RM, RMJ, RPJ and Wu-MAP procedures are not intended to search within $(0,1)$, we convert the points in interval $(0,1)$ by the linear map $6x-3$ to interval $(-3,3)$ and invert the resulting points back to $(0,1)$ for comparison in the same scale. For RM in (\ref{eq:RM-b}), the optimal $a_n=\{nM'(\theta)\}^{-1}$ is used. For RMJ in (\ref{eq:an-bn}), the optimal $\beta=M'(\theta)/\phi(\Phi^{-1}(\alpha))$ and $\tau_1=1$ are used as in \citet{Joseph:04}. For RPJ, set $a_n=n^{-2/3}$ as recommended by \citet{Polyak:Juditsky:92}. For Wu-MAP, set the hyperparameters $\mu_0=0$ and $\tau=\xi=3$ to cover a wide range of priors. For BSA, set $s=17$ to represent a moderate number of sliced subintervals.

Throughout, we set $x_1=0.5$ (corresponding to the starting point zero in $(-3,3)$) and $n=20$ to estimate $\theta$. For $\alpha$ taking values from $0.1,0.2,\ldots,0.9$, we compute the empirical root of mean square (RMSE) of $x_{21}$ over 1,000 replications for every procedure.

\begin{figure}
\centerline{\epsfig{file=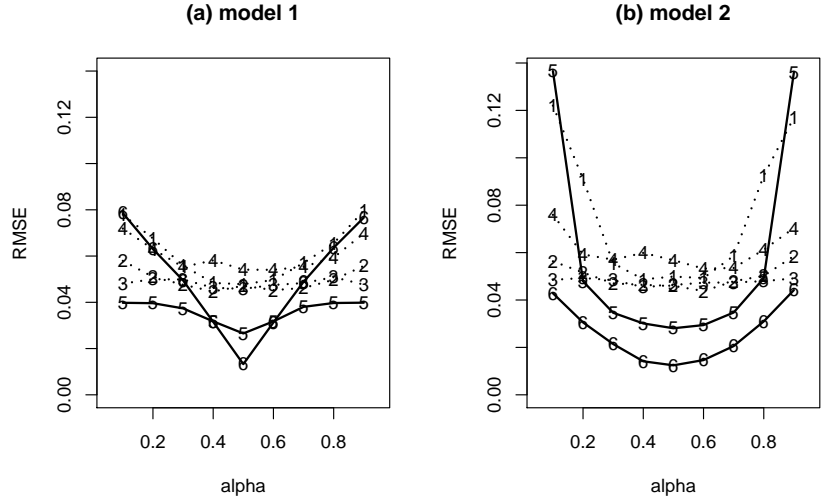, width=4.5in}}
\caption{Empirical RMSEs (over 1,000 replications) of $x_{21}$ obtained by six competing methods (RM `1', RMJ `2', RPJ `3', Wu-MAP `4', BSA-Bayes `5' and BSA-MAP `6') under model 1 (panel a) and model 2 (panel b) for $\alpha=0.1,0.2,\ldots,0.9$, respectively.}\label{fig:rmse_M1M2}
\end{figure}

Figures~\ref{fig:rmse_M1M2} shows the empirical RMSE of $x_{21}$ obtained by the six competing methods. The findings are summarized as follows. (i) Under model 2, the RMJ and RPJ methods perform similarly. Both are superior to the RM and Wu-MAP methods, especially for extreme values of $\alpha$. The proposed method with Bayes estimator has uniform superiority to the RM, RMJ and RPJ methods for $\alpha=0.2,\ldots,0.8$. For $\alpha$ being extreme values as 0.1 or 0.9, the response curve is nearly flat at $\theta$. The performance of BSA-Bayes deteriorates, as expected. While, the proposed method with MAP estimator in this case is the best due to the starting point advantage and its conservatism of movement as pointed out in Example~\ref{exp:BayesvsMAP}. (ii) Under models 3 to 7, the results are similarly to those under model 2. For the sake of space, we defer them in the supplementary material. (iii) Under model 1, the $\alpha$-quantiles of standard normal locates across the search domain. It is seen that the performances of RM, RMJ, RPJ, Wu-MAP are similar to those under model 2. The proposed method with Bayes estimator outperforms the above four methods for all different $\alpha$ values. The RMSE of BSA-MAP has the minimum value for the median estimation and increases in the distance between the root and the starting point which is again because of its conservative movement.

Further simulation shows that the proposed method with other moderate number of subintervals, say $s=15 \sim 25$, yields similar superior result. The implicit stochastic approximation method by (20) of \citet{Toulis:Airoldi:15} was also conducted and found to be much inferior to the RMJ and RPJ methods in the small sample case. The results are omitted.

At last, we want to add that the proposed method requires the uniqueness of $\theta$. When $M'(\theta)$ is very close to zero such as at $M^{-1}(0.9)$, $M^{-1}(0.99)$ or $M^{-1}(0.999)$, the proposed method can perform inferior to the algorithm-based methods RMJ or RPJ. In that case, a hybrid method that uses RMJ or RPJ afer a moderate number of iterations of the proposed method can be used.

\section{Applications}\label{sec:application}
We present two applications of the proposed Bayesian stochastic approximation method for binary responses in this subsection.

\subsection{Search for the root of a monotonic continuous function}\label{sec:root}
For the original problem in (\ref{eq:reg-origin}), first convert $y_n\in\mathbb{R}$ to a response in $(0,1)$ through a sigmoid function $y_n^*=(1+e^{-by_n})^{-1}$, where $b$ is a known scale parameter such that $y_n^*$ spreads well in $(0,1)$. For example, if $y_n$ has a known range in $(-C,C)$ for some $C>0$, we can set $b=3/C$.

Second, approximate $y_n^*$ by a fraction represented by $a$ ones and $q-a$ zeros such that $a/q$ is closest to $y_n^*$ for some integer $q\ge 1$. These $q$ binaries are then treated as independent responses at the same point $x_n$. The minimum value of $q=1$ corresponds to the dichotomization of $y_n$ by its sign. Usually a number as small as $q=3$ is adequate for the approximation.

Based on the generated binary responses, the problem is reduced to search for the median of a distribution. We can then use the proposed method with Bayes estimator in Section~\ref{sec:method}. More specifically, we set $s=5$ for the first ten steps and set $s=9$ for the subsequent steps as used in Section~\ref{sec:simu}.

\begin{figure}[t]
\centerline{\epsfig{file=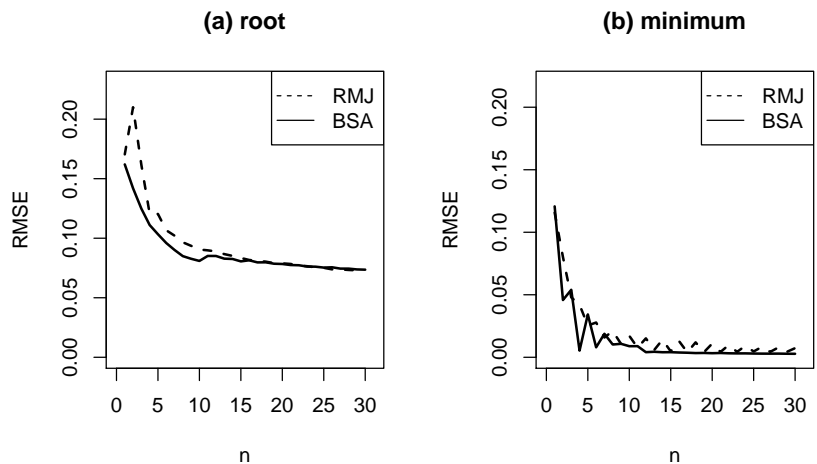,width=4.5in}}\par
\caption{(a) RMSE of $x_n$ for the root up to 30 steps obtained by the proposed method in Application I (in solid line) and by the RMJ procedure (in dotted line), (b) RMSE of the approximated minimum up to 30 steps obtained by the proposed method in Application II (in solid line) and by the RMJ procedure (in dotted line)}\label{fig:rmse_root_min}
\end{figure}

\begin{example}\label{exp:root}
Consider the regression model $y_n=200(x_n-0.3)^3+\varepsilon_n$, where $\varepsilon_n$ is independent standard normal variable. We applied the proposed method above with $b=1$, $q=2$ and $x_1=0.5$. Panel (a) of Figure~\ref{fig:rmse_root_min} shows the empirical RMSEs (over 1,000 replications) of $x_n$ up to 30 steps in comparison with those obtained by applying the (scaled) RMJ procedure (with the same starting point) to the binaries obtained by signs of $y_n$. It is seen that the proposed method dominates the RMJ procedure.
\end{example}

\subsection{Search for a minimum of a convex function}\label{sec:KW}
Suppose that $\varphi(x)$ is a convex function. We seek a sequential design for finding the minimum of $\varphi(x)$ at $\theta$. It is equivalent to find $\theta$ such that $G(\theta)=0$, where $G(x)=\lim_{c\rightarrow 0}\{\varphi(x+c)-\varphi(x-c)\}/(2c)$. The Kiefer--Wolfowitz procedure \citep{Kiefer:Wolfowitz:52} entails the recursion
\[
x_{n+1}=x_n-\frac{\gamma_n\left(y_{n1}-y_{n2}\right)}{c_n},
\]
where $y_{n1}$ and $y_{n2}$ are two independent responses at $x_n+c_n$ and $x_n-c_n$ with mean $\varphi(x_n+c_n)$ and $\varphi(x_n-c_n)$ respectively,
$\gamma_n$ and $c_n$ are two positive constant sequences decreasing to zero and satisfying
$\sum\gamma_n=\infty$, $\sum\gamma_n c_n<\infty$, and $\sum \gamma_n^2c_n^{-2}<\infty$. For example, $\gamma_n=n^{-1}$ and $c_n=n^{-1/3}$ as recommended by \citet{Kiefer:Wolfowitz:52}.

Let $\widetilde y_n=(y_{n1}-y_{n2})/c_n$. We apply the previous procedure in Section~\ref{sec:root} to $(x_n,\widetilde y_n)$ to approximate the root of $G$.

\begin{example}
Consider the regression model $y_n=200(x_n-0.3)^2+\varepsilon_n$, where $\varepsilon_n$ is independent standard normal variable. We conducted a similar comparison using the competing methods in Example~\ref{exp:root} to $\widetilde y_n$ and $c_n$ defined above. Panel (b) of Figure~\ref{fig:rmse_root_min} shows the proposed method outperforms the method based on RMJ procedure in terms of RMSE.
\end{example}

\section{Multi-dimensional extension}\label{sec:extension}
\subsection{Method}
We extend the proposed method for quantile estimation to the multi-dimensional case.

Let $M(\vx)$ be the distribution function of a $p$-dimensional continuous random vector $\vx=(x_1,\ldots,x_p)^\top$ with the domain scaled in the unit hypercube $(0,1]^p$. The goal is to find the generalized multivariate quantile defined by
\[\vtheta=\underset{\{\vx: M(\vx)=\alpha\}}{\argmin}\ U(\vx)\quad\textrm{for}\quad 0<\alpha<1,\]
where $U(\vx)$ is a known function. This is a special case of the notion of generalized multivariate quantiles introduced by \citet{Einmahl:Mason:92}. Like the univariate case,  assume that $\vtheta$ is unique.

The idea of the extension is to use a conditional approach to reduce the problem to univariate case along each coordinate so that the proposed method in Section~\ref{sec:method} can be applied.

First, we introduce some notations. Divide $(0,1]$ equally into $s$ subintervals along each coordinate. For any $\vx\in (0,1]^p$, let $t_{j}=\lceil x_{j}s \rceil$ for $j=1,\ldots,p$ and $\vt=(t_1,\ldots,t_p)^\top$. Then, $\vx$ is uniquely contained in the hypercube
$H(\vx)=\prod_{j=1}^p\left(\frac{t_{j}-1}{s}, \frac{t_{j}}{s}\right]$. Let $\1_p$ denote a vector of $p$ ones and $\ve_a$ denote the $a$th column vector of the $p\times p$ identity matrix. Denote the following $p+1$ vertexes of $H(\vx)$ by
\begin{equation}\label{eq:v}
\vv_0=s^{-1}(\vt-\1_p),\quad \vv_a=\vv_{a-1}+s^{-1}\ve_a,\ a=1,\ldots,p.
\end{equation}
Notice that $\vv_0,\vv_1,\ldots,\vv_p$ are arranged in a helix.

Second, approximate $M$ in $H(\vx_n)$ by the segments of $p$ hyperplanes intersected by the hypercube respectively. The $j$th hyperplane passes through the point $(\vx^{(j)},\alpha)$ with
\[\vx^{(j)}=(x_{n1},\ldots,x_{n,j-1},\theta_j,x_{n,j+1},\ldots,x_{np})^\top\]
and is expressed as
\begin{equation}\label{eq:Fj}
F_j(\vx)=\alpha+\vbeta^\top(\vx-\vx^{(j)}),
\end{equation}
where $\vbeta=(\beta_1,\ldots,\beta_p)^\top$ with $\beta_1,\ldots,\beta_p$ being all positive.

For $a=0,1,\ldots,p$, let $\rho_{a}=F_j(\vv_a)$ and $\vrho=(\rho_0,\ldots,\rho_p)^\top$. Then, by (\ref{eq:v}) and (\ref{eq:Fj}), we have $\rho_{0}<\rho_{1}<\cdots<\rho_{p}$ and the solution of $(\theta_j,\vbeta)$ in $\vrho$ given by
\begin{align}
\beta_a&=\frac{\rho_{a}-\rho_{a-1}}{v_{aa}-v_{a-1,a}}=
s(\rho_{a}-\rho_{a-1}),\quad a=1,\ldots,p,  \label{eq:beta} \\
\theta_j&=v_{0j}+\frac{(v_{jj}-v_{j-1,j})(\alpha-\rho_0)-\sum_{a\ne j}(x_{na}-v_{0a})(\rho_a-\rho_{a-1})}{\rho_j-\rho_{j-1}}. \label{eq:theta-j}
\end{align}
Let $\widetilde\beta_a=\beta_a(v_{aa}-v_{a-1,a})=\rho_a-\rho_{a-1}$ for $a=1,\ldots,p$ and $\widetilde\vbeta=(\widetilde\beta_1,\ldots,\widetilde\beta_p)^\top$. The Jacobian of the transformation from $\vrho$ to $(\theta_j,\widetilde\vbeta)$ is $s\widetilde\beta_j$.

Assume the joint prior of $\vrho$ is uniform with density
\begin{equation}\label{eq:prior-rho-joint}
h(\vrho)=\frac{(p+1)!I(\rho_L<\rho_0<\rho_1<\cdots<\rho_p<\rho_U)}{(\rho_U-\rho_L)^{p+1}}.
\end{equation}
Further denote $\widetilde\vbeta_{-j}=(\widetilde\beta_1,\ldots,\widetilde\beta_{j-1},
\widetilde\beta_{j+1},\widetilde\beta_p)^\top$. By (\ref{eq:beta}), (\ref{eq:theta-j}) and (\ref{eq:prior-rho-joint}), the joint prior of $(\theta_j,\widetilde\vbeta)$ is
\begin{equation}\label{eq:prior-beta-theta}
h(\theta_j,\widetilde\vbeta)=\frac{(p+1)!s
\widetilde\beta_jI\left(0<\widetilde\beta_j<\eta_j(\theta_j,\widetilde\vbeta_{-j}),\ \widetilde\vbeta_{-j}\in \Delta_j\right)}{(\rho_U-\rho_L)^{p+1}},
\end{equation}
where
\begin{align}
\eta_j(\theta_j,\widetilde\vbeta_{-j})&=
\frac{\{\rho_U-\alpha_{1j}(\widetilde\vbeta_{-j})\}I(\theta_j\le \theta_{0j})}{s(v_{pj}-\theta_j)}
+\frac{\{\alpha_{0j}(\widetilde\vbeta_{-j})-\rho_L\}I(\theta_j> \theta_{0j})}{s(\theta_j-v_{0j})}, \notag \\
\alpha_{0j}(\widetilde\vbeta_{-j})&=\alpha+\sum_{a\ne j}\widetilde\beta_as(v_{0a}-x_{na}),\notag \\
\alpha_{1j}(\widetilde\vbeta_{-j})&=\alpha+\sum_{a\ne j}\widetilde\beta_as(v_{pa}-x_{na}),\notag \\
\theta_{0j}(\widetilde\vbeta_{-j})&=
\frac{\{\rho_U-\alpha_{1j}(\widetilde\vbeta_{-j})\}v_{0j}
+\{\alpha_{0j}(\widetilde\vbeta_{-j})-\rho_L\}v_{pj}}{\rho_U-\alpha_{1j}(\widetilde\vbeta_{-j})
+\alpha_{0j}(\widetilde\vbeta_{-j})-\rho_L},\notag \\
\Delta_j&=\left\{\widetilde\beta_a>0\ \textrm{for all}\ a\ne j,\ \alpha_{0j}(\widetilde\vbeta_{-j})>\rho_L,\ \alpha_{1j}(\widetilde\vbeta_{-j})<\rho_U\right\}. \label{eq:simplex}
\end{align}
It is seen that the joint prior distribution of $\widetilde\vbeta_{-j}$ is uniform on the simplex $\Delta_j$ defined by (\ref{eq:simplex}). Then given $\widetilde\vbeta_{-j}$, the conditional distribution of $\theta_j$ after integrating out $\widetilde\beta_j$ and imposing the restriction $0<\theta_j<1$ is
\begin{equation*}\label{eq:h0-j}
h_0(\theta_j\mid\widetilde\vbeta_{-j})=\frac{(p+1)!V_js\eta_j^2(\theta_j,\widetilde\vbeta_{-j})}
{2c_{0j}(\rho_U-\rho_L)^{p+1}},
\end{equation*}
where $V_j$ is the volume of $\Delta_j$ (depending on $\vx_n$) and $c_{0j}$ is the conditional normalization constant (depending on $\widetilde\vbeta_{-j}$).

Alternatively, express
\begin{equation}\label{eq:prior-beta-theta-2}
h(\theta_j,\widetilde\vbeta)=\frac{(p+1)!s
\widetilde\beta_jI\left(\widetilde\beta_j>0, \ell_j(\widetilde\vbeta)<\theta_j<u_j(\widetilde\vbeta),\ \widetilde\vbeta_{-j}\in \Delta_j\right)}{(\rho_U-\rho_L)^{p+1}},
\end{equation}
where
\begin{equation*}\label{eq:Lj-Uj}
\ell_j(\widetilde\vbeta)=v_{pj}-\frac{\rho_U-\alpha_{1j}(\widetilde\vbeta_{-j})}{s\widetilde\beta_j},\quad u_j(\widetilde\vbeta)=v_{0j}+\frac{\alpha_{0j}(\widetilde\vbeta_{-j})-\rho_L}{s\widetilde\beta_j}.
\end{equation*}
The further restriction of $0<\theta_j<1$ which amounts to $0\le\ell_j(\widetilde\vbeta)<u_j(\widetilde\vbeta)\le 1$ requires
$\widetilde\beta_{j0}\le\widetilde\beta_j\le\widetilde\beta_{j1}$ with
\begin{equation}\label{eq:beta-j-0-1}
\widetilde\beta_{j0}=
\max\left\{\frac{\rho_U-\alpha_{1j}}{sv_{pj}},\frac{\alpha_{0j}-\rho_L}{s(1-v_{0j})}\right\},\quad
\widetilde\beta_{j1}=\rho_U-\rho_L-\sum_{a\ne j}\widetilde\beta_a.
\end{equation}

Denote the subsequence of $\vx_n$ contained in $H(\vx_n)$ by $\vx_{i_1},\ldots,\vx_{i_m}$ with $1\le m\le n$. Express the likelihood of $\vx_i\in H(\vx_n)$ as
\begin{equation*}\label{eq:Li-p}
L_i(\theta_j,\widetilde\vbeta)=F_j(\vx_{i})^{y_i}\{1-F(\vx_{i})\}^{1-y_i}=
a_i+b_i(\theta_j)\widetilde\beta_j,
\end{equation*}
where
\begin{align}\label{eq:ai-bi-j}
a_i(\widetilde\vbeta_{-j})&=1-y_i+(2y_i-1)\alpha_i(\widetilde\vbeta_{-j}),\quad
\alpha_i(\widetilde\vbeta_{-j})=\alpha+\sum_{a\ne j}\widetilde\beta_a s(x_{ia}-x_{na}),\\
b_i(\theta_j)&=s(2y_i-1)(x_{ij}-\theta_j);\notag
\end{align}
or as
\[
L_i(\theta_j,\widetilde\vbeta)=
\widetilde{a}_i(\beta_j\mid\widetilde\vbeta_{-j})+\widetilde{b}_i(\widetilde\beta_j)\theta_j,
\]
where
\begin{equation}\label{eq:ai-bi-tilde-j}
\widetilde{a}_i(\widetilde\beta_j\mid\widetilde\vbeta_{-j})=1-y_i+(2y_i-1)\{\alpha_i(\widetilde\vbeta_{-j})+s\widetilde\beta_j x_{ij}\},\quad \widetilde{b}_i(\widetilde\beta_j)=-s(2y_i-1)\widetilde\beta_j.
\end{equation}
It should be noted that unlike the univariate case here $a_i$ depends not only on $(\vx_i, y_i)$ but also the current point $\vx_n$ through $\alpha_i$.

Combining the joint prior of $(\theta_j,\widetilde\vbeta)$ in (\ref{eq:prior-beta-theta}) or (\ref{eq:prior-beta-theta-2}) and the likelihood of the subsequence, we get the posterior distribution of $(\theta_j,\widetilde\vbeta)$ proportion to
$h(\theta_j,\widetilde\vbeta)\prod_{k=1}^mL_{i_k}(\theta_j,\widetilde\vbeta)$.
Given $\widetilde\vbeta_{-j}$, the conditional posterior distributions of $\theta_j$ and $\widetilde\beta_j$ are obtained in the same way as in the univariate case in Sections~\ref{sec:method} and \ref{sec:remark-post-beta} respectively. We summarize the results in the following proposition.

\begin{proposition}\label{prop:post-hmj-gmj}
Assume that the joint prior of $\vrho$ associated with the vertexes of the hypercube $H(\vx_n)$ is uniform with density (\ref{eq:prior-rho-joint}). Then, the conditional posterior distribution of $\theta_j$ restricted in $(0,1)$ is
\begin{equation*}\label{eq:hm-j}
h_m(\theta_j\mid\widetilde\vbeta_{-j})=
\frac{2c_{0j}h_0(\theta_j\mid\widetilde\vbeta_{-j})}{c_{mj}}
\sum_{r=0}^m \frac{d_{m,r}(\theta_j)\eta_j^{r}(\theta_j,\widetilde\vbeta_{-j})}{r+2},
\end{equation*}
where $d_{m,r}$ is defined in (\ref{eq:dm-r}) with $a_i$ and $b_i$ given in (\ref{eq:ai-bi-j}) and $c_{mj}$ is the conditional normalization constant. And the  conditional posterior distribution of $\widetilde\beta_j$ is
\begin{equation*}\label{eq:gm-j}
g_m(\widetilde\beta_j\mid\widetilde\vbeta_{-j})=
\frac{(p+1)!s\widetilde\beta_j}{\widetilde{c}_{mj}(\rho_U-\rho_L)^{p+1}}
\sum_{r=0}^m \frac{d_{m,r}(\widetilde\beta_j\mid\widetilde\vbeta_{-j}) \left\{u_j(\widetilde\vbeta)^{r+1}-\ell_j(\widetilde\vbeta)^{r+1}\right\}}{r+1},
\end{equation*}
where $d_{m,r}$ is defined in the same form as (\ref{eq:dm-r}) with $a_i$ and $b_i$ replaced by
$\widetilde{a}_i(\widetilde\beta_j\mid\widetilde\vbeta_{-j})$ and $\widetilde{b}_i(\widetilde\beta_j)$ in (\ref{eq:ai-bi-tilde-j}) respectively,
and $\widetilde{c}_{mj}$ is the conditional normalization constant over the range $(\widetilde\beta_{j0},\widetilde\beta_{j1})$ given in (\ref{eq:beta-j-0-1}).
\end{proposition}
Proposition~\ref{prop:post-hmj-gmj} reduces to the results in Propositions~\ref{prop:post-hm} and \ref{prop:post-gm} when $p=1$.

The next design point along the $j$th coordinate is then taken to be
\[
\vx_{n+1}^{(j)}=(x_{n1},\ldots,x_{n,j-1},\widetilde\theta_j,x_{n,j+1},\ldots,x_{np})^\top,
\]
where
\[\widetilde\theta_j=\E_{\widetilde\vbeta_{-j}}
\left\{\E_{h_m}\left(\theta_j\mid\widetilde\vbeta_{-j}\right)\right\}\quad \textrm{or}\quad
\E_{\widetilde\vbeta_{-j}}
\left\{\textrm{MAP of }\ h_m\left(\theta_j\mid\widetilde\vbeta_{-j}\right)\right\}.
\]
Since $h_m(\theta_j\mid\widetilde\vbeta_{-j})$ is completely determined, the conditional expectation of $\theta_j$ can be numerically calculated. The expectation with respect to $\widetilde\vbeta_{-j}$ can be approximated by averaging the conditional expectations over finite number of $\widetilde\vbeta_{-j}$ taken uniformly from the simplex. For instance when $p=2$, the simplex $\Delta_j$ for $\widetilde\beta_{-j}$ reduces to $(0,u_{-j})$ where
\begin{equation}\label{eq:upBound-u-j}
u_{-j}=\min\left(\frac{\alpha-\rho_L}{s(x_{n,-j}-v_{0,-j})}, \frac{\rho_U-\alpha}{s(v_{2,-j}-x_{n,-j})}\right),
\end{equation}
$x_{n,-j}$, $v_{0,-j}$ and $v_{2,-j}$ are the other element of $\vx_n$, $\vv_0$ and $\vv_2$ after removing the $j$th element, respectively

At last, we use $U$ to determine the next design point out of the $p$ candidates, i.e.,
\begin{equation}\label{eq:nextPt}
\vx_{n+1}=\vx_{n+1}^{(j^*)} \quad\textrm{with}\quad j^*=\underset{j=1,\ldots,p}\argmin\ U(\vx_{n+1}^{(j)}).
\end{equation}

Note that in the multi-dimensional case, the derivation of the marginal posterior distribution of $\rho_0,\ldots,\rho_p$ is much complicated than the univariate case in Section~\ref{sec:post-rho}. Moreover, there are in fact $p$ different ways to update $\rho_L$ (or $\rho_U$) in (\ref{eq:prior-rho-joint}) depending on the coincidence of $\vv_0$ (or $\vv_p$) with {\em one} vertex of some neighboring hypercube. So in the following numerical study, we simply fix $\rho_L=0$ and $\rho_U=1$ for all hypercubes and let the data inside the hypercube learn the posterior distribution of the interested parameter.

\subsection{Numerical illustration}
We illustrate the proposed method by a few examples. Consider the following three models:
\begin{align*}
M_8(x_1,x_2)&=\Phi(6x_1-3,6x_2-3,0),\\
M_9(x_1,x_2)&=\Phi(6x_1-3,6x_2-3,0.8),\\
M_{10}(x_1,x_2)&=\Phi(6x_1-3,6x_2-3,-0.8),
\end{align*}
where $x_1,x_2\in (0,1)$ and $\Phi(z_1,z_2,\rho)$ is the distribution function of bivariate normal variables with zero means, unit marginal variances and correlation coefficient $\rho$.

We use the same two-stage procedure with respect to the choice of $s$ as for the univariate case in Section~\ref{sec:simu}. For illustration, we set the starting point $\vx_1=(0.6,0.6)^\top$ for all cases and recommend using MAP estimator for $\alpha=0.25$ to be conservative. The uniform distribution for $\widetilde\beta_{-j}$ over $(0,u_{-j})$ in (\ref{eq:upBound-u-j}) is approximated by a discrete uniform distribution over $\{iu_{-j}/8: i=1,\ldots,7\}$.

In these examples, by symmetry  we have $\theta_1=\theta_2$ and the determination for the next point in (\ref{eq:nextPt}) can be modified as $j^*=\argmin_{j=1,2}|x_{n,-j}-\widetilde{\theta}_j|$, i.e. to choose a point that is closer to the diagonal line $x_1=x_2$.

\begin{figure}[h]
\centerline{\epsfig{file=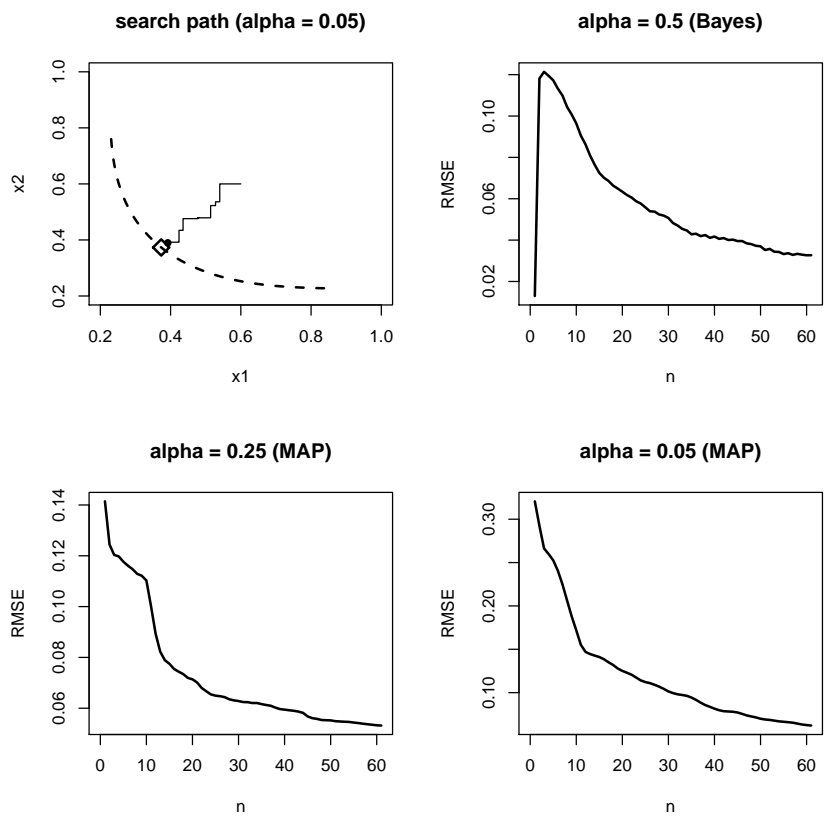,width=4.5in}}\par
\caption{(a) one search path under $M_8$ with $\alpha=0.05$, where the dotted curve represents the set $M_8^{-1}(0.05)$ and $\vtheta$ is indicated by `$\diamondsuit$'; (b), (c) and (d) the empirical RMSE (over 1,000 replications) of $\vx_n$ obtained by the proposed method with different estimator in parenthesis for $\alpha=0.5$, 0.25, and 0.05, respectively.}\label{fig:h1p2}
\end{figure}

Panel (a) of Figure~\ref{fig:h1p2} presents a single search path under $M_8$ with $\alpha=0.05$, where the dotted curve is the solution set of $M_8^{-1}(0.05)$ and $\vtheta=(0.3733,0.3733)^\top$ is indicated by `$\diamondsuit$'. Panels (b), (c) and (d) of Figure~\ref{fig:h1p2} show the empirical RMSE (over 1,000 replications) of $\vx_n$ up to 60 steps obtained by the proposed method using different estimators (in parenthesis). The convergence of the procedure is clear. For the case with $\alpha=0.5$, the small value of RMSE at the first few steps is due to the starting point.

The results for models 9 and 10 are similar and hence omitted.

\section{Conclusion and discussion}\label{sec:disc}
The proposed Bayesian stochastic approximation method uses an adaptive local model and  yields a recursive updating scheme in terms of the posterior distribution in stead of the estimate itself. It has the advantage of successively utilizing the information of the neighboring points to improve the estimation efficiency, thus reduces the variation or uncertainty carried by a single point. However, there remain several questions unsettled. First, the asymptotic behavior of the procedure in both univariate and multivariate cases is not fully understood. Second, the refined prior in both univariate case and multi-dimensional case is worth further investigation. Third, more efficient algorithm is desired, especially for multi-dimensional situation, where information about the posterior distribution of $\widetilde\beta$ can be used.

Because of the rich and broad applications of stochastic approximation, we anticipate new explorations of the proposed method in interactions with different techniques in many fields that mentioned at the beginning of the article.

R package is provided in the supplementary material.

\section*{Acknowledgements}
 The research is supported by the National Natural Science Foundation of China
(grant 11271134) and the 111 Project (B14019) of Chinese Ministry of Education.

\section*{Appendix}
\begin{figure}[h]
\centerline{\epsfig{file=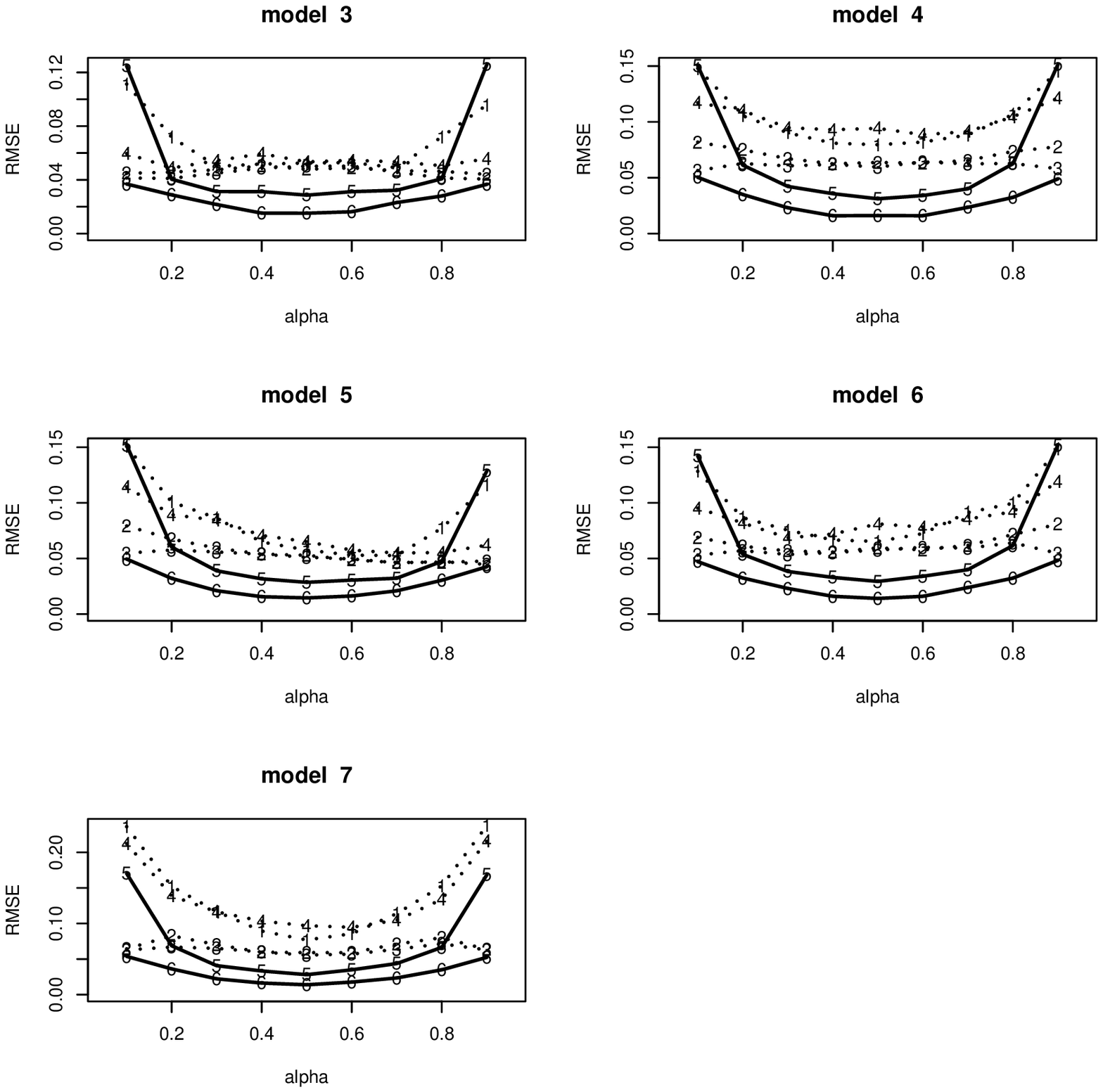,width=5.0in}}
\caption{Empirical RMSEs (over 1,000 replications) of $x_{21}$ obtained by six competing methods (RM `1', RMJ `2', RPJ `3', Wu-MAP `4', BSA-Bayes `5' and BSA-MAP `6') under models 3$\sim$7 for $\alpha=0.1,0.2,\ldots,0.9$, respectively.}\label{fig:rmse_M34567}
\end{figure}

Figure~\ref{fig:rmse_M34567} shows the empirical RMSE of $x_{21}$ obtained by the six competing methods under models 3 to 7. The results are similar to those obtained under model~2.


\end{document}